\documentclass{article}
\usepackage{psfig}
\begin{document}
% \draft

% \preprint{}

\title{Opening A New Window to the Early Universe}

\author{Eric Hivon\footnote{Electronic address: \tt
     efh@ipac.caltech.edu}  ~and Marc
     Kamionkowski\footnote{Electronic address: \tt
     kamion@tapir.caltech.edu} \\
{California Institute of Technology, 
     Pasadena, CA 91125, USA}}

\maketitle
%\pacs{}
%]

% \begin{centering}
%  Science vol 298, p 1349, 15 Nov 2002
% \end{centering}

% \vskip 1cm

The big news at the recent Cosmo '02 workshop in
Chicago \cite{cosmo02} was the announcement of the
first detection of polarization in the cosmic microwave
background (CMB), the 2.726 K radiation left over from the big
bang \cite{dasi}.  

In 1968, Rees predicted that the CMB must be polarized if it is a relic from the
early universe\cite{rees}. Ever since, astronomers have sought observational
evidence. The race for detection heated up after precise measurements of
temperature fluctuations  \cite{experiments} provided increased
confidence in our ability to understand the CMB. The new
discovery, reported by the Degree Angular Scale Interferometer (DASI)
collaboration, not only confirms our theoretical grasp of the CMB, 
but also opens a whole new window to the early universe. 

Early-universe cosmology merges the search for new laws of
fundamental physics, beyond the standard model of particle
physics and Einstein's gravity, with the search to understand
the origin and evolution of the universe.  The mean thermal
energies of particles in the primordial soup that filled the
universe microseconds after a big bang greatly exceed those
accessible with our most powerful terrestrial particle
accelerators.  The early universe thus provides a test bed for
new ideas for ultra high-energy physics -- if it has left a trace in today's
universe, the big bang's cosmic debris. Fortunately, a truly pristine 
cosmological relic exists: the CMB.

To a very good approximation, the temperature of the CMB
radiation is the same in all directions in the sky.  However, at
the level of 1 part in $10^5$, there are small variations.  
The CMB radiation was emitted $\sim14$ 
billion years ago when electrons and nuclei first combined to form atoms, at a
time when the universe was only $\sim$ 400,000 years old.  
Thus, the angular temperature variations reflect variations in properties (such as density,
pressure, temperature and  velocity) of the primordial universe.

\begin{figure}[htbp]
\centerline{\psfig{file=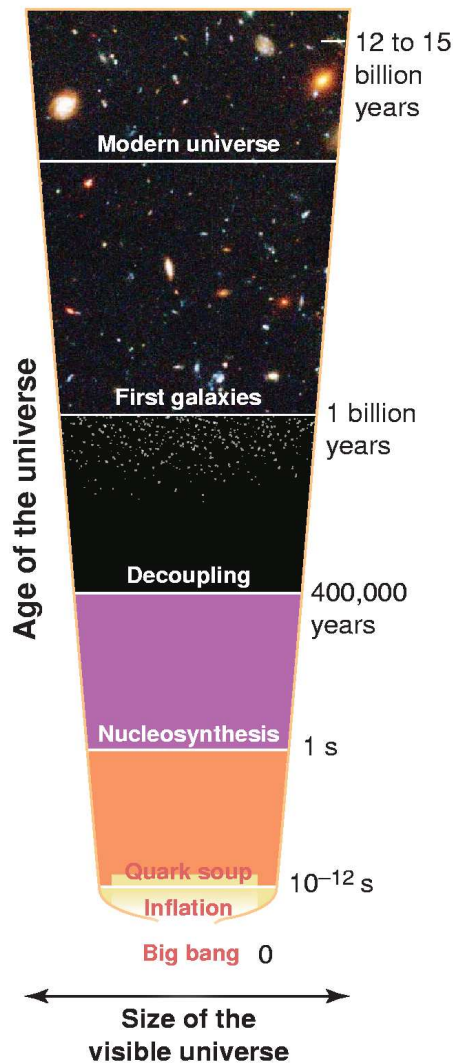,width=2.5in}}
\medskip
\caption{{\bf From smooth to structured.}  The big bang may have been followed by a period
     of rapid inflation, during which the resulting ``soup'' of particles
     coalesced into nucleons and lighter elements.  Matter and
     radiation eventually became decoupled, the former
     gravitationally clumping into the structure of the modern
     universe and the latter yielding the microwave background we see 
     today.  The seeds from which galaxies grew should be apparent in 
     the variations in the radiation background.
}
\label{fig:history}
\end{figure}

The temperature patterns
at the CMB surface of last scatter were probably inscribed
even earlier, just fractions of a microsecond after the
big bang (see Fig.~\ref{fig:history}).  Particle theories suggest
that in the extreme temperature that existed then, gravity may have briefly
become a repulsive, rather than attractive, force.  The enormously accelerated
expansion during the ensuing period of ``inflation'' can explain the remarkable
smoothness of the CMB and produce the primordial mass
inhomogeneities imprinted in the CMB temperature.

Existing CMB temperature maps allow the temperature
power spectrum, which quantifies the size distribution of hot
and cold spots, to be determined.  Comparisons with
predictions of inflation models for primordial inhomogeneities then provide
constraints for several cosmological parameters (such as the mass 
density, the geometry of the universe and its expansion rate).  
Moreover, the oscillatory pattern seen in the CMB power spectrum \cite{SZ}
confirms that the primordial inhomogeneities are consistent with inflation. 

The CMB polarization contains yet more cosmological
data than that provided by the temperature maps alone.  Most
light is unpolarized (the orientation of the
oscillating electric field that makes up the electromagnetic
wave is random).  But light can also be linearly polarized (the field is more
likely to oscillate in a given direction).  In the CMB, the polarization
indicates a direction at the surface of last scatter.
However, the
polarization amplitude is very small -- just $\sim$ 7\% of the
temperature-fluctuation amplitude for the polarization from
primordial inhomogeneities.  
Inflationary models make many predictions for the statistical properties of the
polarization \cite{KamKos}.

The current DASI results (see the figure)
are not yet nearly precise enough to test the inflationary
predictions fully, but they are a dramatic first step.
They detect the polarization with high confidence ($5\sigma$), and the
measured amplitude is consistent with that expected.  

Far more will be learned with
more precise polarization maps.
First, the polarization will provide much more precise velocity maps because it
is due primarily to the velocity at the
surface of last scatter. In contrast, the temperature pattern is due to a
combination of the mass inhomogeneity and velocity. 
Second, the polarization provides a test for inflation theories, which predict a unique
polarization pattern \cite{waves}.  
Third, polarization might map the mass
distribution in the more recent universe through the effects of
weak gravitational lensing \cite{lensing}. The galaxies between us and the surface
from which the CMB radiation was emitted induce a
gravitational bending of light that leads to an identifiable
distortion to the CMB polarization pattern. Finally, polarization
with large coherence patches is generated by rescattering of CMB radiation from
intergalactic debris produced by the onset of star formation.

DASI has ended a 34-year quest to detect the CMB polarization, sounding the
starting gun for a new race to peer further back in time, with more
precision than ever before.  Many more CMB polarization experiments are in
progress or planned. NASA's recently launched Microwave Anisotropy Probe
(MAP) \cite{map} should detect the large-angle
polarization induced by early star formation. This should be
followed by increasingly precise ground and balloon
experiments leading to the Launch of the European Space Agency's
Planck satellite \cite{Planck} in 2007. If the recent past
is any indication, studies of the CMB will continue to advance cosmology, even
after Planck.

\begin{figure}[htbp]
\centerline{\psfig{file=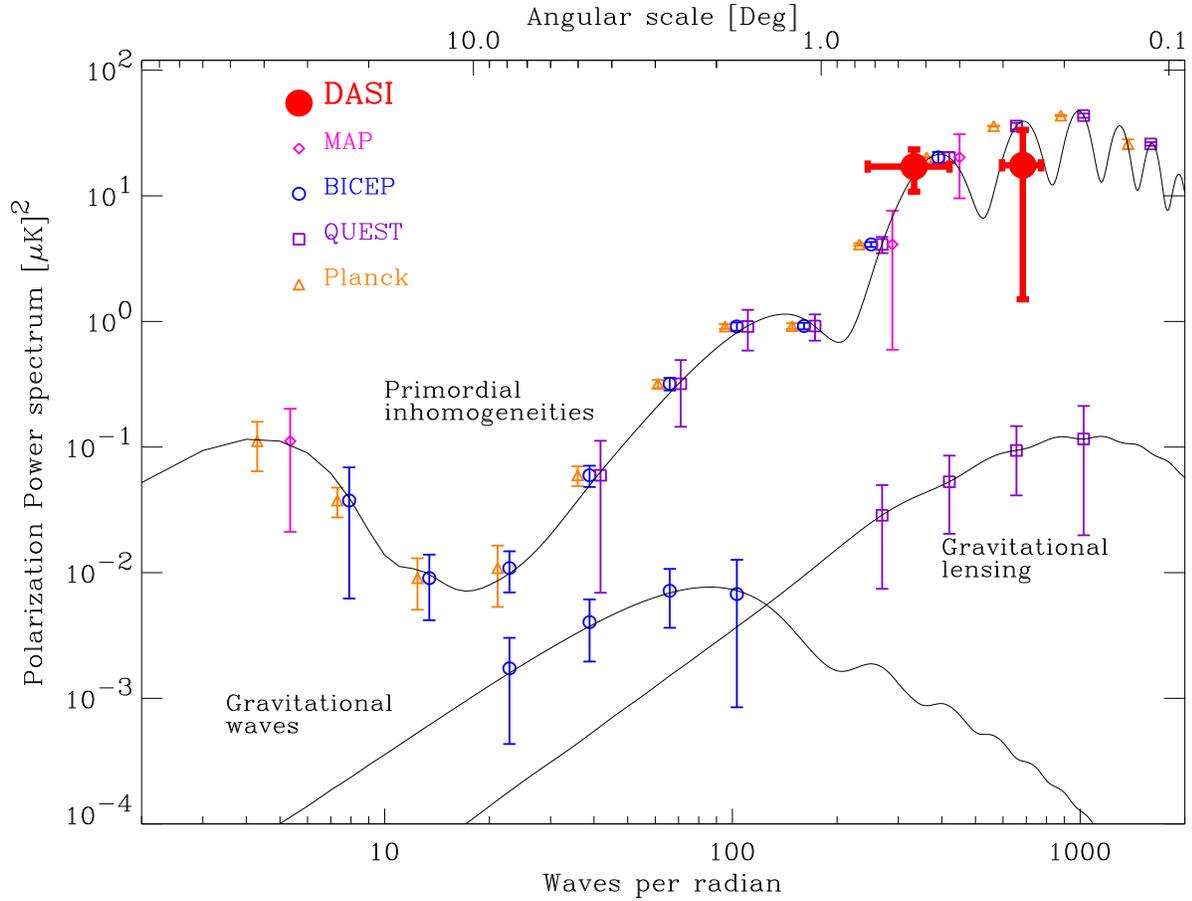,width=6.5in,angle={90}}}
\medskip
\caption{{\bf Current and future polarization data.}  
	The polarization power spectrum determines the correlation of
	polarization over patches of sizes indicated on the top axis. (Top
curve) Prediction for the polarization from primordial inhomogeneities produced
by inflation. The large-angle bump in this curve is the enhancement from early
star formation. (Lower curves) Inflationary gravitational-wave and
gravitational-lensing signals. These can be distinguished from the larger
mass-inhomogeneity signal with geometric properties of the polarization. DASI
data points are shown in red. Future experiments will go beyond DASI in
sensitivity to detect some of the other signals. We show the data points that
experimentalists hope to achieve with some of these new experiments \cite{future}.}
\label{fig:experiments}
\end{figure}

% \begin{references}

\end{document}